\documentclass[prl,twocolumn,showpacs]{revtex4-1}

\usepackage[utf8x]{inputenc}
\usepackage{amsmath,amssymb,amsfonts,subfigure,braket,color,dsfont}
\usepackage[english]{babel}
\usepackage[dvips]{graphicx}
\usepackage{sistyle}

\usepackage{subfig}

\DeclareMathOperator{\tp}{\otimes}
\DeclareMathOperator{\trace}{Tr}
\DeclareMathOperator{\Var}{Var}

\newcommand{\abs}[1]{\left| #1 \right|}

\newcommand{\id}{\mathds{1}}

\newcommand{\R}{\mathbb R}

\begin{document}

\title{The Fisher information and the quantum Cram\'er-Rao sensitivity limit of continuous measurements}
\author{S\o ren Gammelmark}
\author{Klaus M{\o}lmer}
\email{moelmer@phys.au.dk}
\affiliation{Department of Physics and Astronomy, Aarhus University, Ny
  Munkegade 120, DK-8000 Aarhus C, Denmark.}

\date{\today}

\begin{abstract}
Precision measurements with quantum systems rely on our ability to trace the differences between experimental signals to variations in unknown physical parameters. In this Letter we derive the Fisher information and the ensuing Cram\'{e}r-Rao sensitivity limit for parameter estimation by continuous measurements on an open quantum system. We illustrate our theory by application to resonance fluorescence from a laser driven two-state atom and we show that photon counting and homodyne detection records yield different sensitivity to the atomic parameters, while none of them exceed our general result.
\end{abstract}


\maketitle

\nocite{Gill2000State,Gill2013From,Yuen1973Multipleparameter,Belavkin1976Generalized}

Quantum systems find wide applications in high precision measurements, e.g., as clocks and as probes of the strength of  perturbations and of inertial effects. For the situation where an initially prepared quantum system is measured after being subject to an unknown interaction, much theoretical effort has been devoted to identify which are the ideal initial states for such experiments and by which kind of measurement does one optimally distinguish among close candidate values of the quantity probed \cite{Giovannetti2004QuantumEnhanced}.
In this Letter, we derive the quantum sensitivity limit for a different situation where continuous measurements are performed on the radiation emitted over time by an open quantum system. Quantum trajectory analyses \cite{carmichael1993open,Dalibard1992,Wiseman2010Quantum} have been applied to simulate how the interplay of random measurement outcomes and quantum measurement back action gradually filters the candidate values for system parameters \cite{Mabuchi1996Dynamical,Gambetta2001State,Tsang2011,Gammelmark2013Bayesian,Kiilerich2014}. While the achievement of such filters depend on the detection scheme applied, we present in this Letter a new, general theory which fundamentally limits how properties of open quantum systems, are revealed by the continuous interaction with their environment.

Our analysis applies for example to conventional fluorescence detection where the radiation emitted spontaneously on a laser driven atomic two-level transition is a function of the detuning, driving field strength and the atomic decay rate which can therefore be determined, e.g., by fitting the mean fluorescence intensity at different driving frequencies to a Lorenzian frequency profile. It is easy to understand that this fit improves with the total number of detected photons, and thus with accumulation time. The fluctuations in the time dependent fluorescence signal, however, also contribute important information, since following each detection event, the atom jumps to the ground state and is subsequently excited by the laser field. Unlike the mean fluorescence intensity which is power broadened and which saturates at high laser driving power, the distribution of time intervals between photo detection events is an oscillatory function and it allows, by a Bayesian analysis, a   precise discrimination between even large amplitudes of the driving field \cite{Gammelmark2013Bayesian,Kiilerich2014}.

\begin{figure}
  \centering
 \includegraphics{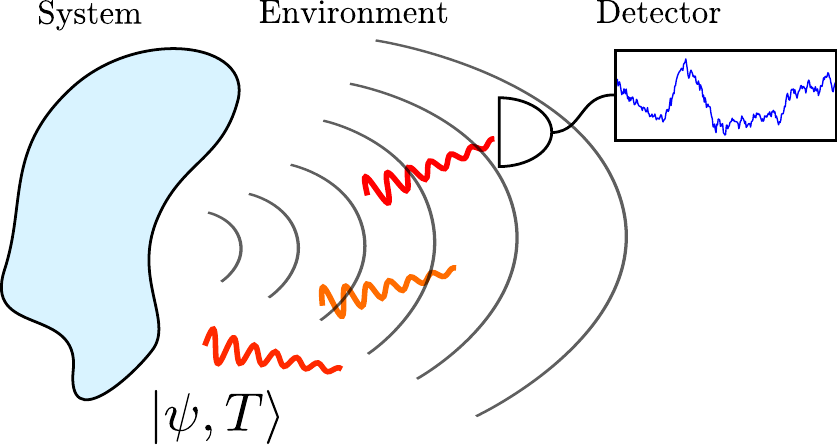}
 \caption{(Color online) A quantum system interacts with the quantized radiation field, the environment, through emission of photons. Under continuous probing until time $T$ the emitter quantum state evolves in a stochastic manner, while in the absence of observation, the full system and environment may at time $T$ be described by a pure quantum state $|\psi,T\rangle$.}
\label{fig:schematic}
\end{figure}

While we aim to account for the information available by continuous measurement of the emitted radiation during a time interval $[0,T]$ , we note that such measurement signals can be equivalently obtained from correlations in the freely propagating field occupying regions of space at different distances from the emitter, i.e., from a joint measurement on the quantum state $|\psi,T\rangle$ of the system and the environment at the final time, see Fig.~\ref{fig:schematic}. The theoretical parameter estimation sensitivity associated with time dependent field measurement records is therefore identical to the estimation sensitivity for the corresponding final state measurement on the tensor product space of the system and the environment. We can hence formally apply the theory by Braunstein and Caves, \cite{Braunstein1994Statistical}, which determines the resolution limit for the determination of an unknown parameter $\theta$ through measurements on a quantum system in a pure state which is a function of $\theta$, see also \cite{Yuen1973Multipleparameter,Belavkin1976Generalized,Holevo1982,Gill2000State}.

Since infinitely many modes are occupied, and since the number of emitted quanta grows with time, the full quantum state of the emitter and the quantized field is not tractable in a practical calculation. As we shall show in the following, however, the sensitivity limit given by Braunstein and Caves, can be easily calculated, as long as the system-environment interaction permits application of the Born-Markov approximation. This includes to an excellent approximation the emission of light by laser driven atoms, but also other system-environment situations, such as microwave emission in circuit QED systems \cite{cqed}, are represented well by our analysis. The Born-Markov approximation ascertains that we can discretize the time evolution such that the emitter effectively interacts with independent environment degrees of freedom in each new time interval $[t_i,t_i+\delta t]$. For atom-light interaction, these degrees of freedom correspond to the quantized radiation field modes occupying shells at different propagation distances from the emitter, cf. Fig.~\ref{fig:schematic}. The total emitter and environment Hilbert space at time $T = N\delta t$ is hence decomposed as the tensor product of the state space of the light emitting system and $N$ radiation field Hilbert spaces, where a unitary operator $U_{t_i}$  acts on the emitter and on the corresponding environment sub-system in each time interval.

By associating with each environment subspace an $M$-dimensional Hilbert space with states $\ket{m_i}$, $m=0, \ldots M-1$, and assuming that each subspace is initialized in a definite state $\ket{0_i}$ prior to interaction with the emitter at $t_i$, we ensure the Markovian behaviour
of the system-environment interaction, and we establish an analogy with the work by Gu\c{t}\u{a} on system identification in quantum Markov chains \cite{Guta2011Fisher}. The derivation of results later in this Letter specific to the continuous time master equations may thus be viewed as an implementation of the more general quantum Markov chain framework, addressed in \cite{Guta2011Fisher}, see also \cite{Hayashi2008,Gill2013From}.

After the system-environment interaction the combined state at time $T$ can be written
\begin{eqnarray} \label{eq:tensorproduct}
\ket{\psi; T} = U_{t_{N-1}} \ldots U_{t_{0}}\ket{\psi_S^0}\tp \ket{0_{N-1}, \ldots 0_0}\nonumber \\
= \sum_{m_0, \ldots m_{N-1}} \Omega_{m_{N-1}} \ldots \Omega_{m_{0}} \ket{\psi_S^0} \tp \ket{m_{N-1}, \ldots m_0}
\end{eqnarray}
where $\ket{\psi_S^0}$ is the initial state of the small system. The operators $\Omega_{m_i}$ identify the action on the emitter at times $t_i$ associated with the transfer of the field sub-system from $\ket{0_i}$ to $\ket{m_i}$: In a basis $\{|k_S\rangle\}$ for the emitter $\Omega_{m_i}$ has the matrix representation $\braket{k_S|\Omega_{m_i}|k_S'} \equiv \braket{k_S, m|U_{t_i}| k_S', 0}$.

A projective measurement performed on the emitted field with outcome $m_i$ projects the environment sub-system on the definite eigenstate $|m_i\rangle$ with $0 \leq m_i < M$ and thus causes a  measurement back action on the emitter given by the corresponding operator $\Omega_{m_i}$. These operators thus constitute {\it measurement effect operators} \cite{Wiseman2010Quantum}, and unitarity of $U_{t_i}$ ensures that they form positive operator-valued measures (POVM), i.e., $\sum_m \Omega_{m_i}^\dagger \Omega_{m_i} = \id$. To a particular time dependent outcome signal $m_0, \ldots m_{N-1}$ corresponds a single term in the sum in Eq.(\ref{eq:tensorproduct}), which constitutes the stochastic wave function or quantum trajectory $|\psi_S^T\rangle_c \propto \Omega_{m_{N-1}} \ldots \Omega_{m_0} \ket{\psi_S^0}$ of the emitter conditioned on the measurement record \cite{Dalibard1992,carmichael1993open,Wiseman2010Quantum}.

The unitary operators $U_{t_i}$, and hence the measurement effect operators $\Omega_{m_i}$, depend on the parameters $\theta$ that we want to estimate by observation of the signal emitted by the quantum system. Evaluating the trace over a complete basis for the environment, one obtains the reduced system density matrix, which evolves according to the $\theta$-dependent linear map $\rho_{t_i} = \sum_{m=0}^{M-1} \Omega_m(\theta) \rho_{t_{i-1}} \Omega_m^\dagger(\theta)$. For an infinitesimal time step $\delta t$, the evolution assumes the  Lindblad form master equation ($\hbar=1$)
\begin{multline}
\frac{d\rho}{dt} =  \mathcal{L}_\theta(\rho) = (\sum_m \Omega_m(\theta) \rho_{t} \Omega_m^\dagger(\theta) - \rho(t))/\delta t \\
 = \frac{1}{i}[H,\rho] + \sum_c (L_c\rho L_c^\dagger - \frac{1}{2}(L_c^\dagger L_c\rho + \rho L_c^\dagger L_c). \label{eq:lindblad}
\end{multline}
In passing to the second line in (\ref{eq:lindblad}), we have introduced a different summation index $c$ to indicate that environment state projections may be combined into a different number of relaxation terms in the reduced system master equation. One may for example associate different directions of emitted photons with the same final ground state of the atom. A Lindblad master equation does not uniquely specify the measurement effect operators, but it may suggest a simple form of no-jump and jump dynamics \cite{Dalibard1992}, corresponding to null-measurements and single photon counts in the case of resonance fluorescence. Thus, with a system Hamiltonian $H$, we may have $\Omega_0 = \id - i (H - \frac{i}{2} \sum_{c=1}^C L_c^\dagger L_c) \delta t$ and $\Omega_c = L_c \sqrt{\delta t}$, $1 \leq c \leq C$.

%Our theory is not restriced to  time-homogeneous measurement processes, but let us for simplicity assume that the unitary operators $U_{t_i}$, while acting non-trivially on different Hilbert spaces, all have the same matrix elements.

For a quantum system which is prepared in a pure quantum state $\ket{\psi;\theta}$ that depends on a parameter $\theta$, Braunstein and Caves \cite{Braunstein1994Statistical} identified the general measurements whose outcome data yield the best parameter sensitivity, and they identified the corresponding state dependent quantum Fisher information $I(\theta)$ with the Bures metric on the quantum state space. An asymptotically large number $K$ of independent measurements with outcome data probabilities that are conditioned on an unknown parameter $\theta$ permits a Bayesian estimate of $\theta$ with a variance given by the Cram\'{e}r-Rao bound, $\Var(\hat\theta) \geq \frac{1}{K I(\theta)}$. For the estimation of a vector $\theta \in \R^n$ of $n$ real unknown parameters, $\theta^{\alpha}$, the Cram\'{e}r-Rao bound correspondingly limits the estimation by a covariance matrix $[K I(\theta)]^{-1}$ for the, possibly correlated, errors, where the quantum Fisher information matrix can be similarly identified with the Bures metric tensor, see, e.g., \cite{Zanardi2007}, and reads,
\begin{align} \label{eq:QuantumFisher}
  I_{\alpha\beta}(\theta) = 4\Re(\braket{\partial_\alpha \psi|\partial_\beta \psi} - \braket{\partial_\alpha\psi |\psi;\theta}\braket{\psi;\theta|\partial_\beta\psi}).
\end{align}
In (\ref{eq:QuantumFisher}), $\Re$ denotes the real part and $\ket{\partial_\alpha\psi} \equiv \partial\ket{\psi;\theta}/\partial\theta^\alpha$.
An equivalent  expression for the quantum Fisher Information matrix involves the variation of the overlap $\langle\psi;\theta_1|\psi;\theta_2 \rangle$ of states depending on two values of the vector of variables $\theta_1,\theta_2 \in \R^n$,
\begin{equation}
I_{\alpha\beta}(\theta) = 4\partial^1_\alpha \partial^2_\beta (\log \abs{\braket{\psi;\theta_1|\psi;\theta_2} }) \lvert_{\theta_1 = \theta_2 = \theta}, \label{eq:FisherInnerProduct}
\end{equation}
where $\partial^{1(2)}_\gamma$ is defined as the derivative with respect to the $\gamma$-component of $\theta_1$ ($\theta_2$).

These expressions for the Fisher information apply when we insert the joint state of the system and environment (\ref{eq:tensorproduct}). But, that state is not available in practice, and it is a main result of the present Letter that we do not need to know the full pure quantum state to evaluate the overlap and the derivatives in Eqs.(\ref{eq:QuantumFisher},\ref{eq:FisherInnerProduct}). From Eq.(\ref{eq:tensorproduct}), we observe that the inner product $\braket{\psi; T, \theta_2|\psi; T, \theta_1}$ can be written as $\trace_{Sys,Env}(\ket{\psi; T, \theta_1}\bra{\psi; T, \theta_2})$ which can in turn be written $\trace_{Sys}(\rho_{\theta_1, \theta_2}(T))$, where the action of the operators $\Omega_{m_{N-1}}(\theta_1) \ldots \Omega_{m_{0}}(\theta_1)$ from the left, and $\Omega^\dagger_{m_{0}}(\theta_2) \ldots \Omega^\dagger_{m_{N-1}}(\theta_2)$ from the right, is accumulated in the generalized, reduced density matrix $\rho_{\theta_1, \theta_2}(T)$. Due to the similarity with the first line in (\ref{eq:lindblad}), we notice that   $\rho_{\theta_1, \theta_2}(t)$ can be defined and calculated as the solution to the generalized master equation
\begin{eqnarray} \label{lindblad12}
\frac{d\rho}{dt}\equiv \mathcal{L}_{\theta_1,\theta_2}(\rho),
\end{eqnarray}
where $\mathcal L_{\theta_1, \theta_2}(\rho) = (\sum_m \Omega_m(\theta_1) \rho \Omega_m^\dagger(\theta_2) - \rho)/\delta t$.

Setting $\theta_1=\theta_2$ in (\ref{lindblad12}), we recover the usual trace preserving master equation, while for different $\theta_1$ and $\theta_2$, (\ref{lindblad12}) yields the non-trivial value of $\braket{\psi; T, \theta_2|\psi; T, \theta_1}$, that we need to calculate the Fisher information. For a quantum emitter of Hilbert space dimension $L$, the master equation involves only the solution of $L^2$ coupled differential equations, which can be readily done if $L$ is not too large. Solving the generalized master equation (\ref{lindblad12}) for $(\theta_1,\theta_2)$ in a neighborhood of $(\theta,\theta)$ we may numerically determine the derivative in (\ref{eq:FisherInnerProduct}).

If the system Hamiltonian and the coupling to the environment are not explicitly time dependent, a unique stationary eigenstate of the usual Lindblad master equation (\ref{eq:lindblad}) obeys, $\mathcal L_\theta(\rho_s) = 0$, and $\mathcal L_{\theta_1, \theta_2}$ has an eigenvalue $\lambda_s(\theta_1, \theta_2)$ which in a small neighbourhood around $(\theta, \theta)$ is smoothly connected to the vanishing eigenvalue of $\mathcal L_\theta$. Assuming that all other eigenvalues of $\mathcal L_\theta$, and hence of $\mathcal L_{\theta_1, \theta_2}$, have finite (negative) real part, the overlap can, for times $T$ much longer than the corresponding relaxation times of the emitter, be approximated by a single term
\begin{align}
  \trace_{Sys}(\rho_{\theta_1, \theta_2}(T)) \sim e^{T \lambda_s(\theta_1,\theta_2)} c_s(\theta_1, \theta_2),
\end{align}
where $c_s(\theta_1, \theta_2)$ is the expansion coefficient of the initial system state ${\ket{\psi_S^0}}{\bra{\psi_S^0}}$ on the eigen-matrix of $\mathcal L_{\theta_1,\theta_2}$ with eigenvalue $\lambda_s$.
Inserting this into (\ref{eq:FisherInnerProduct}) we get
\begin{align}
  I_{\alpha\beta} \sim 4T \partial_\alpha^1 \partial_\beta^2 \Re(\lambda_s(\theta_1, \theta_2))|_{\theta_1=\theta_2=\theta} + O(1). \label{eq:ContinousQuantumFisher}
\end{align}
Rather than relying on numerical differentiation, we can use perturbation theory around $(\theta_1, \theta_2) = (\theta, \theta)$ in the original master equation (\ref{eq:lindblad}). This leads to an expression for $\partial_\alpha^1\partial_\beta^2 \lambda_s$ involving operator expectation values determined in the steady state $\rho_s$ \cite{SupplementaryMaterial}.

Signal contributions separated by more than the emitter correlation time, are qualitatively independent, and hence we may regard the full signal as a number of independent contributions which is proportional to the total data acquisition time, while the parameter sensitivity further invokes the dependence of the dynamics on the parameter in question. The result (\ref{eq:ContinousQuantumFisher}) confirms this qualitative expectation: the Fisher information in the emitted signal is, indeed, proportional with time and we have provided a method to calculate the constant of proportionality.
The master equation (\ref{eq:lindblad}) is invariant under ($\theta$-independent) unitary basis transformations on the environment sub-systems. Our expression for the Fisher information therefore does not depend on the choice of environment basis, and the Cram\'er-Rao bound thus limits the estimation capability of any conceivable time dependent field measurement.

\begin{figure}
\includegraphics[width=\columnwidth]{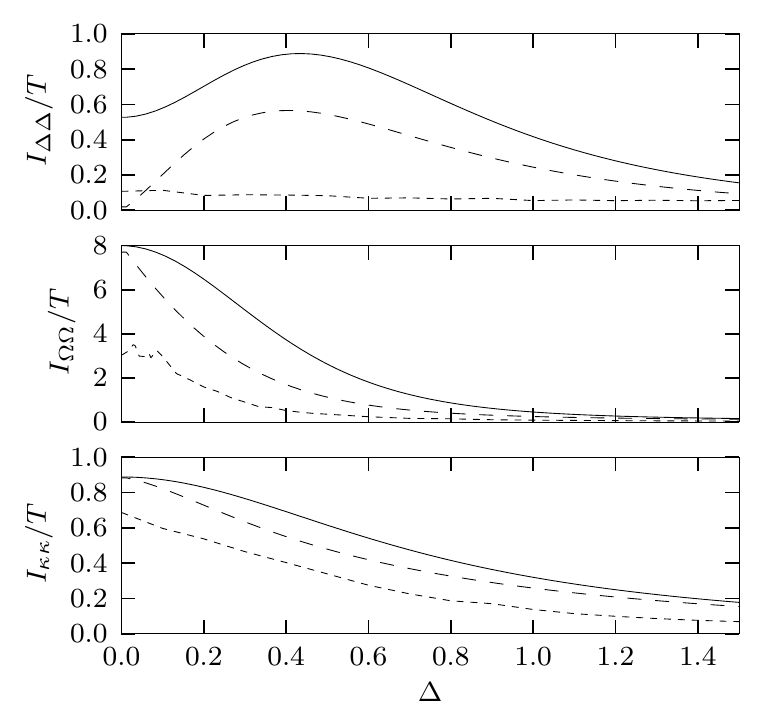}
\caption{Fisher information, divided by the total measurement time, for the probing of light emission by a two-level atom. The upper panel shows the Fisher information for estimation of the detuning $\Delta$, the middle(lower) panel shows the Fisher information for estimation of the Rabi frequency $\Omega$ (decay rate $\kappa$). Within each plot,
the solid line shows the quantum Fisher information, which bounds any measurement strategy, while the long (short)  dashed lines show the Fisher information for photon counting (homodyne detection).}
\label{fig:FisherTwoLevel}
\end{figure}

Let us illustrate the application of  our formalism by considering a two-state atom with a ground state $\ket g$ and excited state $\ket e$ driven by a laser detuned from resonance by $\Delta$ and with a Rabi-frequency $\Omega$. The excited state decays by fluorescence emission with a rate $\kappa$, given in the usual way by the coherent coupling strength to the quantized radiation field and the density of field  modes.

Regardless of the kind of measurement performed on the fluorescence signal, the reduced system master equation  is given by (\ref{eq:lindblad}) with the Hamiltonian $H = \Delta \ket e \bra e + \Omega/2 (\ket e \bra g + \ket g \bra e)$ and a single Lindblad (jump) operator $L = \sqrt{\kappa} \ket g \bra e$.
We assume that $\Delta$, $\Omega$ and $\kappa$ all are in dimensionless units, e.g., relative to
a reference frequency standard.

Let us first assume that both the Rabi frequency and the decay rate are known. The quantum Fisher information for estimation of the detuning $\Delta$ can then be determined by the methods described above, and its slope with respect to time is shown as a function of the detuning by the solid line in the upper panel in Fig.~\ref{fig:FisherTwoLevel} (assuming $\Omega = 1$, $\kappa = 1/2$). For reference, the Fisher information can be computed by a Bayesian likelihood analysis for counting
\cite{Gammelmark2013Bayesian, Kiilerich2014} and homodyne detection \cite{Gammelmark2013Bayesian}, see also \cite{Mabuchi1996Dynamical,Gambetta2001State}, with the results shown as the long and short-dashed lines in the figure. We observe that the two detection schemes offer different degrees of resolution: Near a vanishing detuning, the fluorescence intensity has a maximum, and photon counting is not sensitive to small detuning changes, while phase sensitive, homodyne detection yields a finite Fisher information. For larger detuning, however, the counting signal yields more information than homodyne detection, and both stay below the general limit.

In the second panel we present the quantum Fisher information for estimation of the Rabi frequency $\Omega$, in the vicinity of $\Omega=1$, as a function of the known detuning, and assuming $\kappa=1/2$. In the third panel we present the quantum Fisher information for estimation of the decay rate $\kappa$, in the vicinity of $\kappa=1/2$, as a function of the known detuning, and assuming $\Omega=1$.
At zero detuning, the counting signal exhausts the information available about the Rabi frequency and about the decay rate, and for all detunings, counting yields better sensitivity to both parameters than homodyne detection.
The three panels only display the diagonal elements of the Fisher information matrix, $I_{\Delta \Delta},\ I_{\Omega \Omega}$ and $I_{\kappa \kappa}$, and they yield the limit for how well one may estimate any of the parameters if the other two are known, while the full Fisher information matrix is needed to estimate the errors if two or all three parameters are unknown.

If the system Hamiltonian or the system-environment interaction are time dependent, or if the system is interrogated for only a finite time, the system density matrix does not converge to a stationary state, and the above eigenvalue analysis does not apply directly. We can still, however, determine the Fisher information by solving the time dependent, generalized master equation (5) with different parameter values $(\theta_1,\theta_2)$ and subsequently determine the derivatives numerically. Alternatively, we can apply linear response theory and obtain an expression which involves a two time correlation function, which may in turn be evaluated by use of the quantum regression theorem. Such a calculation, which is briefly indicated in the Supplementary Material \cite{SupplementaryMaterial}, is a reflection of the fact that while the emitted field is formally eliminated in the reduced system master equation (\ref{eq:lindblad}), the fluctuating field observables, subject to our detection, can in the Heisenberg picture be expressed in terms of the system dipole operator \cite{Kimble1976Theory}. The master equation and the quantum regression theorem thus yield equations of motion for both mean values and two- and multi-time correlation functions of the emitted field \cite{Lax1968Quantum}.

Notably, with the possibility to treat a time dependent system-environment interaction, we have also the possibility to describe a system coupled to different meter degrees of freedom, and hence we can also account for the Fisher information associated with measurements carried out over time, and possibly at the end of the experiment, on the small quantum system itself.

If the reduced system density matrix populates a subspace that does not couple to the environment and if it occupies a superposition state that evolves with a frequency proportional to the unknown system parameters, a final measurement on the system rather than on the emitted radiation yields a parameter resolution of $1/T$ and hence a Fisher information $I \propto T^2$ rather than $I \propto T$. This case violates our eigenvalue analysis because of the existence of more than one eigenvalue with vanishing real part, while a more careful treatment indeed yields the different scaling of the Fisher information with time. For a detailed discussion and an example of this particular situation, see \cite{Guta2011Fisher}. Note that the two-time integral in calculations using linear response theory and the quantum regression theorem  scales with $T$, if the system correlations have finite lifetime, while it scales quadratically with $T$ if they are undamped \cite{SupplementaryMaterial}. A similar change of the Fisher information between linear and quadratic  scaling was observed in a recent analysis of weak value probing \cite{Walmsley2013}. We believe that this result reflects how probing itself gives rise to a damping term in the reduced system master equation. In the limit of vanishing probe interaction, the system evolves freely, and the parameter resolution scales with $1/T$.

%  (not depending on the parameters $\theta$) will leave these results unchanged.
%It is easily seen that the maps $\cleft{\alpha}$, $\cright{\beta}$, $\mathcal C_{\alpha\beta}$ and $\mathcal C$ are all invariant under the transformation for a unitary matrix $V$.
%
%We can obtain a different characterization of the Fisher information useful for asymptotic times $T$.

In summary, we have derived general expressions for the limit by which the monitoring of an open quantum system described by a Lindblad master equation can yield precise information about unknown system parameters.  For the example of a light emitting two-level atom, Fig.~\ref{fig:FisherTwoLevel} compares the general sensitivity with the ones calculated by a Bayesian likelihood analysis for photon counting and homodyne detection. This quantative analysis on the one hand confirms the validity of our analysis, and on the other hand it raises a natural question concerning which measurement schemes achieve the highest sensitivity limit. Further candidate measurement schemes that may be worth investigating include weak field homodyne detection \cite{Wamsley}, hybrid counting and homodyne detection \cite{Orozco}, and adaptive measurements \cite{Wisemanadapt, Mabuchiadapt}.

From a more general perspective, our work may also be the starting point for investigations of how bounds from quantum information theory and complementarity arguments may limit the ability to actually achieve the Cram\'er-Rao bound for certain combinations of parameters, e.g., associated with non-commuting interaction Hamiltonians, and for restricted types of measurement. Due to the formal similarity of the quantum trajectory terms in (1) with Matrix Product States \cite{Verstraete2008a,Ostlund1995}, and to their association with dynamical phase transitions and criticality \cite{Lesanovsky2013}, we also expect that our method may lead to identification of systems and parameter regimes with sensitivity scaling different from $1/\sqrt{T}$ and $1/T$, cf, the sensitivity of quantum many body systems near phase transitions \cite{Zanardi2007,Tsang2013}.

We acknowledge support from the Villum Foundation and from the Aarhus University Research Foundation.

\end{document}